# Label-Free Non-Contact Structural and Functional Vascular Imaging using Photon Absorption Remote Sensing

James A. Tummon Simmons[1,2], Sarah J. Werezak[1], Benjamin R. Ecclestone[1,2], James E. D. Tweel[1,2], Hager Gaouda[1,2], Parsin Haji Reza[1*]

[1] PhotoMedicine Labs, University of Waterloo, 200 University Ave W, Waterloo, ON N2L 3G1, Canada.
[2] illumiSonics Inc., 22 King Street South, Suite 300, Waterloo, ON N2J 1N8, Canada.
*Corresponding author: phajireza@uwaterloo.ca

**Abstract** – Vascular imaging is critical for understanding human health and disease. Most established non-contact and label-free optical techniques capture predominantly structural information about vasculature. However, in many pathologies, functional changes often precede visible morphological changes. This limits the ability of established modalities to prevent negative patient outcomes. In this study, a new *in vivo* Photon Absorption Remote Sensing (PARS) microscope is proposed. PARS enables the label-free non-contact structural and functional imaging of vascular structures and their microenvironment. PARS aims to capture the dominant light matter interactions surrounding an absorption event including both non-radiative and radiative relaxation. System performance is demonstrated through wide field of view *in vivo* imaging of vascular contrast in both mouse ear and chicken embryo. Additionally, video rate imaging of chicken embryo capillaries shows first feasibility for blood flow measurement using PARS. This work represents a promising step for vascular imaging techniques where there is significant demand for a method of non-contact label-free functional imaging.

## I. Introduction

Vascular imaging plays an essential role in detailed investigations of vascular mechanisms for critical diseases including cancer and stroke, as well as major diseases of the eye such as diabetic retinopathy and age-related macular degeneration [1]–[4]. Ideally vascular imaging tools would capture structural and morphological context, accurate measurement of functional parameters such as blood oxygenation and blood flow, and ultimately identification of specific molecular compounds. Additionally, each of these contrasts should be characterized across a wide range of scale, including down to the smallest capillary, which is best achieved by optical methods.

Optical vascular imaging techniques derive contrast from the interactions of light with matter, typically from scattering and absorption. Scattering based vascular imaging techniques include light reflectometry, confocal microscopy, and optical coherence tomography (OCT) [5], [6]. These techniques capture detailed structural and morphological information including vessel location, diameters, orientation, and details about surrounding tissue including neuron structure and tissue thicknesses. Functional extensions of these techniques have been developed which include hyperspectral reflectometry and visible OCT for blood oxygenation [7], [8], as well as doppler OCT for blood flow [9], and OCT angiography for detailed vasculature mapping [10], [11]. Although pervasive and successful, these techniques face intrinsically limited chromophore specificity, and subsequently limited accuracy for functional parameters such as blood oxygenation due to low absorption sensitivity [12].

Alternatively, absorption-based imaging techniques typically observe the radiative or non-radiative relaxation of molecules after an excitation event. In absorption microscopes, exceptional molecular specificity can be achieved through careful consideration of excitation wavelength. Vascular imaging modalities which rely on radiative contrast include fluorescence microscopy [13], [14], multi-photon fluorescence microscopy [15], multi-harmonic generation microscopy [16], and stimulated emission microscopy [17]. These techniques capture great vascular contrast, with fluorescent microscopy techniques acting as the backbone of fundamental biological research. However, while radiative emissions can be captured from either endogenous or exogenous chromophores, many of these techniques use exogenous labelling to provide vascular contrast. This makes them particularly powerful for derived animal models and *ex-vivo* samples, but consequently limits applicability to human imaging. This is a result of key biomolecules for vascular imaging, namely hemoglobin, having very low quantum yield, meaning most absorbed energy is released through fast non-radiative transitions limiting endogenous fluorescence contrast [18], [19].

As a result, non-radiative absorption microscopes have seen success in vascular imaging especially in photoacoustic microscopy (PAM) [20], [21] with some preliminary work in photothermal methods [22], [23]. PAM and its derivatives has demonstrated exceptional volumetric imaging of blood vessels, and accurate measurement of functional parameters such as blood oxygenation and blood flow [24]–[26]. However, due to a contact-based transducer, PAM is limited in applications where non-contact operation is essential including in eye, brain, and wound imaging [27].

Photon absorption remote sensing (PARS) microscopy is a novel non-contact, label-free imaging technique which aims to simultaneously capture the dominant light matter interactions around photon absorption events [28], [29]. PARS uses a co-focused pulsed excitation and continuous wave detection laser to induce and observe excitation events, capturing both non-radiative and radiative relaxation pathways. Where the deposited heat and pressure from non-radiative relaxation cause modulation of local optical properties which are captured as non-radiative contrast through intensity modulations of the probe laser. While radiative contrast is captured through measuring the intensity of the emitted photons during radiative relaxation. PARS has evolved from photoacoustic remote sensing reported by Haji Reza et al. [30], [31] which has demonstrated significant promise in vascular imaging in dermatological, ophthalmic, and neurologic applications [30]–[34] using only non-radiative contrast. However, through adding sensitivity to radiated photons, the presented *in vivo* PARS improves the molecular specificity through capturing the total absorption [28]. Altogether, PARS microscopy represents a powerful alternative for applications where contact is prohibitive, such as in ophthalmic, neurological, or surgical applications.

This work demonstrates the first application of PARS to the *in vivo* imaging of vascularized tissues. To improve repeatability, we present a novel system control and alignment interface which involved the creation of new non-radiative extraction methods to mitigate image artifacts and noise as well as extensive use of the complementary contrasts of excitation and detection scattering for alignment. These improvements are all demonstrated through large field of view PARS imaging of vasculature in chicken embryo and mouse ear, where the first videos of individually resolved flowing red blood cells using PARS are demonstrated. Altogether, these key system improvements in contrast and repeatability demonstrate a maturing microscopy system capable of capturing detailed images of vascular structure and function while remaining entirely non-contact and label-free.

## II. Methods

### A. in vivo PARS Optical System Architecture

The *in vivo* PARS imaging system architecture is depicted in Figure 1. This system integrates a pulsed visible excitation laser, and a continuous wave near infrared detection laser. The selected excitation laser is a 532 nm visible 1.5 ns pulsed ytterbium-doped fiber laser (GLPN-16-1-10-M, IPG Photonics), and the detection laser is an 830 nm near-infrared (NIR) super-luminescent diode (SLD) laser (SLD830S-A20, Thorlabs). The output of the excitation laser is condensed through a Galilean beam expander (GBE02-A, Thorlabs), before passing through a 70:30 power-based beam splitter (BS023, Thorlabs) used for excitation backscattering sensitivity. The excitation beam then passes through a dichroic mirror used to separate the radiative emissions circulator (VCIR-3-830-S-L-10-FA, Ascentta). The output of this fiber circulator is then collimated (C40APC-B, Thorlabs) and optically combined with the excitation path using an 805 nm returning from the sample (DMSP605, Thorlabs) before being optically combined with the detection laser. The detection laser fiber is fitted into a manual paddle polarization controller (FPC560, Thorlabs), before being coupled into a

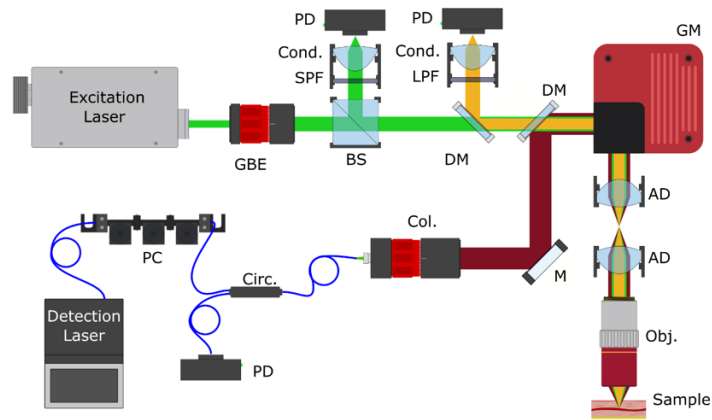

**Figure 1.** *In vivo* PARS microscope simplified system diagram depicting detection (red), excitation (green), and radiative (yellow) pathways. System component abbreviation definitions are as follows: *AD* achromatic doublet, *BS* beam splitter, *Circ.* circulator, *Col.* collimator, *Cond.* condenser lens, *DM* dichroic mirror, *GBE* Galilean beam expander, *GM* galvanometer mirrors, *LPF* long pass filter, *M* mirror, *Obj.* objective, *PC* polarization controller, *PD* photodetector, *SPF* short pass filter.

Label-Free Non-Contact Structural and Functional Vascular Imaging using Photon Absorption Remote Sensing

short pass dichroic mirror (DMSP805, Thorlabs). The optically combined detection and excitation paths are then scanned in two dimensions using a galvanometer mirror pair (GVS012/M, Thorlabs). The pivot point of the galvanometer mirror pair is then imaged onto the pivot point of the objective (Plan Apo NIR 20X, Mitutoyo), using a telecentric pair composed of two cemented achromatic doublets (AC254-050-AB-ML, Thorlabs).

Three photodiodes are then used to capture optical interactions with the sample. The back-scattered intensity of the detection light returning from the sample is retransmitted through the same pathway as described in the forward propagation. Upon arriving back in the circulator, it is redirected onto an avalanche photodiode (APD130A, Thorlabs) where the intensity of back-scattered light is captured before, during, and after excitation. Similarly, the back-scattered intensity of the excitation light is retransmitted back through the system where 70% of returning light is reflected from the power-based beam-splitter where it is focused by a condenser lens (ACL25416U-A, Thorlabs) onto a photodiode (PDB425A, Thorlabs). Finally, the radiative emissions returning through the system predominately pass through the 805 nm short pass dichroic mirror (DMSP805, Thorlabs), and are isolated through reflection off a 605 nm short pass dichroic mirror (DMSP605, Thorlabs). These emissions are then focused using a condenser lens (ACL25416U-A, Thorlabs) onto an avalanche photodiode (APD130A, Thorlabs).

### B. Signal Capture and Image Formation

*Image Capture Workflow:* To capture an image, the galvanometer mirrors were used to optically scan the sample in a two-dimensional raster scan pattern. The fast axis scan pattern was a triangle wave, while the slow axis scan pattern was made of a custom ascending and descending staircase waveform. The pulsed excitation laser trigger was used as the digital system sampling trigger, where the three information carrying photodiode signals (detection scattering, excitation scattering, and radiative emissions) in addition to the galvanometer mirror analog positions were sampled around each excitation pulse. Specifically, ~4 μs of each of these signals were sampled at a frequency of 250MHz using multiple digitizers in parallel to capture PARS contrasts simultaneously (ATS9626, AlazarTech). Both laser pulse repetition rate, and mirror speeds were carefully controlled to sample the frame respecting Nyquist sampling theorem.

*PARS System Signal Extraction:* The *in vivo* PARS imaging system captures four unique optical contrasts around each excitation event including detection scattering, excitation scattering, non-radiative absorption, and radiative absorption. Both the non-radiative absorption, and detection scattering are extracted from the time resolved intensity measurements of the back-scattered detection light. Two forms of non-signal extraction are described including zero-order baseline correction, Figure 2 (a), and first-order baseline correction, Figure 2 (b). Zero order baseline corrected extraction is described in Equation (1) where the non-radiative signal intensity ($I_{NR}$) is extracted as the percentage change of the difference in the integrated detection intensity ($S_{det}$) before (A) and during (B) excitation. This extraction was previously leveraged to generate detailed absorption contrast in slides [28], [29].

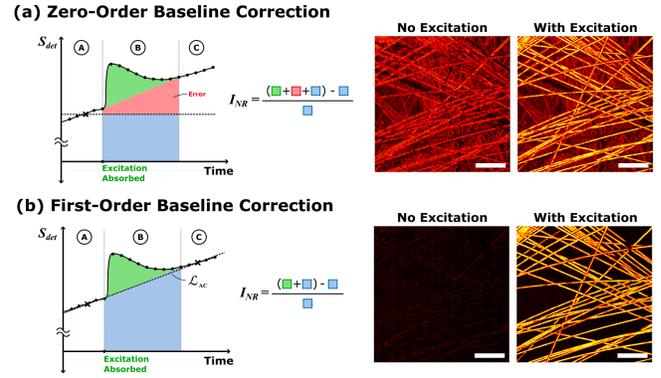

**Figure 2.** Overview of non-radiative signal contrast extraction. (a) Shows the zero-order baseline correction extraction. Includes a graphical depiction of a single time resolved detection intensity signal with non-radiative modulation (green), baseline (blue), and error (red) regions (on left). A graphic summarizing Equation (1) (middle). The non-radiative contrast extracted from an image of carbon fibre both with, and without excitation (right). The same three subsections are shown for the first order extraction in section (b).

$$I_{NR} = \frac{\sum^B S_{det} - \sum^A S_{det}}{\sum^A S_{det}} \quad (1)$$

While the zero-order baseline correction is sufficient for slide scanning systems[28], [29], significant error can be introduced when optically scanning at high speeds. As a result, for the *in vivo* PARS microscope the non-radiative contrast is calculated using a first-order baseline corrected extraction. Where to account for significant bias in estimated non-radiative signal intensity, a linear fit ($\mathcal{L}_{AC}$) is made based on scattering intensity ($S_{det}$) before (A) and after (C) excitation. This allows for the non-radiative intensity ($I_{NR}$) to be more accurately calculated avoiding edge artifacts introduced due to the higher speed optical scanning.

$$I_{NR} = \frac{\sum^B S_{det} - \sum^B \mathcal{L}_{AC}}{\sum^B \mathcal{L}_{AC}} \quad (2)$$

Finally, all of excitation scattering, radiative absorption, and analog mirror position are extracted based on integrating total signal energy. Altogether this produces four unique optical contrasts, and two position feedback signals.

Label-Free Non-Contact Structural and Functional Vascular Imaging using Photon Absorption Remote Sensing

*Image Parameters:* Two scan patterns were selected to achieve the chosen optical field of view with the highest speed while allowing sufficient relaxation time for the sample. Two frame capture sizes were used in this work including a slower, larger 512μm x 512 μm frame which ran at a framerate of ~0.25Hz with a ~82 kHz excitation trigger, and a 64μm x 64μm frame running at a framerate of ~20Hz with a ~77 kHz excitation trigger.

*Image Formation:* To generate an image from collected data, the fast and slow mirror signals are segmented into lines and then populated discretely into the image frame. Next, using the full frame a circular shift operation is performed in order to correct for mirror position lag removing left-right scan artifacts. For larger mosaics, images were collected at discrete positions with a known offset and overlap by leveraging electromechanical stages. Distinct sub-images were then aligned based on maximizing two-dimensional correlation between adjacent frame overlaps. Overall images were then intensity corrected based on non-linear levelling methods from recent work [29].In order to improve image visualization images were both median and gaussian filtered.

### C. Alignment Interface and Digital System

To support rapid alignment and acquisitions during experimentation a PARS graphical user interface (GUI) was developed to provide detailed real time feedback. The newly developed PARS GUI leverages a highly parallelized multi-threaded architecture built in C++ to manage the high data volumes of >500 MB/s continuously. Through real-time processing it displays essential visualizations of signal time domains, spatial contrast profiles, as well as line-by-line reconstruction of images all enabled by a state-of-the-art immediate mode GUI library (Dear ImGui). Additionally, control of most digital system components were integrated into the GUI including the arbitrary waveform generators used to provide analog input to the mirrors, as well as the excitation laser trigger. This new interface has completely replaced and extended the functionalities of previously presented methods [35].

### D. Animal Handling and Laser Safety

*Chicken embryo models.* Chicken embryo models were cultured and maintained in accordance with University of Waterloo Health Research Ethics Committee (Protocol ID: 44703). Briefly, fertilized eggs (White Leghorn) were incubated in a consumer egg hatcher for 3 days before being carefully transferred to custom chicken embryo holders and then placed in an incubator (GFQ Manufacturing, SKU 1502W) which maintains a suitable temperature (~37 °C) and humidity (>70%). Chicken embryo models were then used for imaging >10 days post incubation.

*Mouse models.* This study presents results from Charles River SKH1 Hairless Mice. All experimental procedures using *mus* models were carried out in accordance with University of Waterloo Health Research Ethics Committee (Protocol ID: 44703). For imaging, animals were anesthetized using a 5% isoflurane/oxygen mixture for induction, before mounting in a custom animal holder, at which a ~1.5% isoflurane/oxygen mixture was used to maintain suitable levels of anesthesia through observation of breathing rate, physical responses, etc. Additionally, the animal's body temperature was maintained at ~37 °C using an infrared thermal heating pad, and the ear was treated with Nair to remove extraneous hairs prior to imaging (Nair, Church & Dwight Co., Inc.).

*Laser safety.* Imaging powers in all animal models were maintained to <4mJ/$cm^2$ which are well within skin safety limits of 20mJ/$cm^2$ [30].

## III. Results and Discussion

The presented *in vivo* PARS system offers many improvements over past iterations of PARS vascular imaging systems. These include developments in optical, digital, and software systems as well as signal processing methods. Most significantly this imaging system embodiment incorporates sensitivities to detection, and excitation scattering, as well as non-radiative absorption, and radiative emissions from the sample. This offers a wealth of information which proves critical for performance of the system both in terms of valuable contrast, and repeatability.

### A. Signal Extraction and PARS Imaging

*Signal Extraction:* As described in Figure 2, significant improvements were made to non-radiative signal extraction with the implementation of first-order baseline removal. As can be seen in Figure 2 (a, b; right) the newly derived extraction method significantly outperforms previous methods. Here the non-radiative signal is better isolated from the changing detection scattering levels, which may introduce significant noise on sharp changes where the slope of the scattering is steep. This was introduced as a major challenge for the *in vivo* PARS imaging system due to the high optical scan speeds. This had not been a problem for previous work, as all other PARS system iterations have been entirely stage scanning, meaning the scattering intensity does not typically change significantly within one time domain making the zero-order approximation sufficient.

*Mirror Scanning:* There have been significant improvements to image FOV primarily led by improvements to scan patterns. As described in the methods, custom raster scans are generated to scan the sample in a raster pattern using an ascending and descending staircase waveform. This enables better mirror lag compensation, while maintaining a particular imaging sampling density across the scan area. This enables both significantly larger single frame acquisition areas (512x512µm) than previously presented, as well as higher imaging acquisition rates at small FOVs (~64x64µm, ~20Hz) while retaining correct spatial sampling density and avoiding image artifacts.

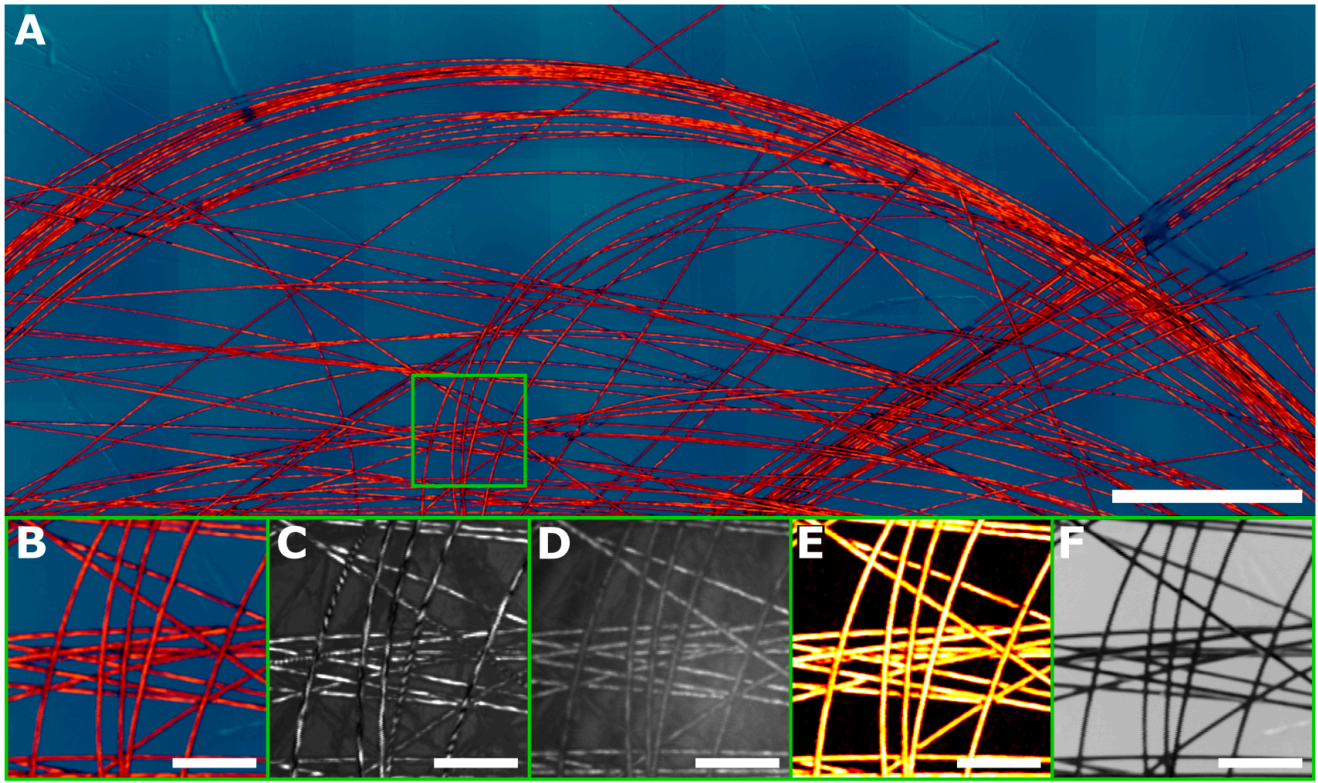

**Figure 3.** PARS images of alignment target composed of carbon fibre bundles placed on a fluorescent alignment target submerged in water. (a) Total absorption PARS image (scale bar: 500 µm). (b) Zoom in of total absorption contrast from highlighted area in A (scale bar: 100 µm). (c) Detection scattering contrast highlighting surface structure of carbon fibres. (d) Excitation scattering contrast which is directly complementary to detection scattering. (e) Non-radiative absorption contrast highlighting high absorption, and non-radiative relaxation of carbon fibres. (f) Radiative absorption contrast highlighting radiation from fluorescent target with a long depth of focus.

*PARS Alignment and Acquisition GUI:* The real time interface has enabled significant repeatability improvements through providing the user additional information about system parameters including relative detection and excitation lateral and axial focal arrangement. These kinds of essential alignment metrics including intensity profiles across individual line scans, and detailed line-by-line reconstruction of our image contrasts enables faster and more justified alignment of PARS imaging systems. Additionally, the stability improvements offered by these changes in aggregate have enabled more advanced mosaicking resulting significantly larger fields of view across a range of samples through multiple image acquisitions.

### B. Imaging of PARS Alignment Targets

Performance of the *in vivo* PARS imaging system was characterized by imaging a carbon fibre alignment target which consists of 7 µm carbon fibres placed on top of a highly fluorescent alignment target which were then both submerged in water. The captured PARS contrasts for a ~3.5x1.5mm mosaic image of the alignment target is shown in Figure 3. As shown, the four captured PARS contrasts provide highly complementary information. Figure 3 (a) shows a composite image depicting a PARS total absorption image highlighting the non-radiative and radiative relaxation which is occurring in the sample. This provides good context for two materials of well-known quantum yields where carbon fibres exhibit almost complete lack of radiative emissions, and the lower fluorescent target exhibits almost no non-radiative relaxation. This can be seen in more detail in Figure 3 (b) which shows many crossing fibres and the consistent measured absorption signal. The detection and excitation scattering images, provided in Figure 3 (c) and (d) respectively, show similar structural images of the surfaces of the carbon fibres. The detection scattering, shown in Figure 3 (c), demonstrates both a higher sensitivity and resulting contrast based on the confocal conditioning imposed by the fibre coupling back into the circulator. This accounts for the differences seen in the excitation scattering, shown in Figure 3 (d), which includes a smoother, longer depth of field image of the same carbon fibres due to the free space optical detection pathway. Finally, the two direct absorption measurements are shown in Figure 3 (e) and (f). Figure 3 (e) shows the non-radiative signal where the carbon fibres can be seen to provide predominantly non-radiative relaxation absorption contrast. Figure 3 (f) highlights the long depth of field radiative contrast captured by the PARS imaging system. Alternatively, the non-radiative contrast, provides higher depth sectioning as a result of the confocal detection configuration as shown in Figure 3 (a) and (e). This means that in the combined PARS visualization we can see great structural context provided by the combination of a long depth of focus radiative contrast and a more highly depth-sectioned non-radiative contrast. This PARS alignment target was typically used to assess the performance of the system and optimize both axial and lateral alignment of detection and excitation beams. Real time feedback enabled through the live PARS alignment interface allowed for consistent alignment of system performance before moving to more the following more complex samples.

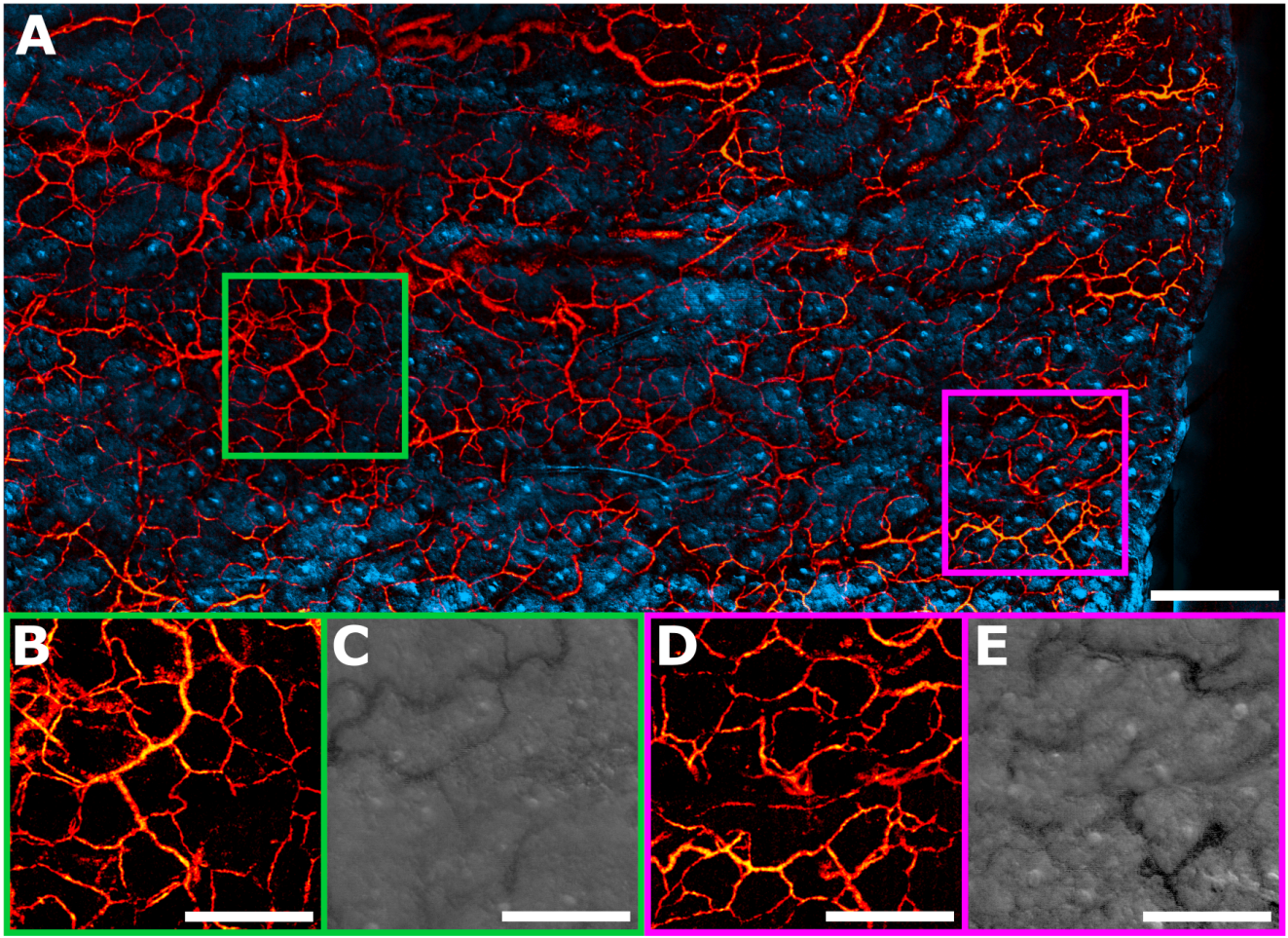

**Figure 4.** PARS images of *in vivo* mouse ear across ~5x2.5mm. (a) Total absorption PARS image of mouse ear highlight complementary contrast (scale bar: 400µm). (b, c) Zoom in of non-radiative (b) and radiative (c) contrast from green highlighted area in A (scale bar: 200µm). (d, e) Zoom in of non-radiative (d) and radiative (e) contrast from pink highlighted area in A (scale bar: 200µm).

### C. PARS Imaging of Mouse Ear

The PARS imaging system was then used to capture detailed images of *in vivo* nude mouse ear which can be seen displayed in Figure 4. The full extent of the ~5x2.5mm field of view image can be seen in Figure 4 (a) through a total absorption PARS image. From this visualization the great complementary absorption contrast offered from observing both the non-radiative and radiative contrasts simultaneously is abundantly clear. Across the ear detailed structures ranging from layers of vasculature contrast to local structural components including cartilage matrices, and sebaceous glands can be seen. This is emphasized in the total absorption contrast in Figure 4 (a), where the vasculature can actually be clearly seen wrapping around and supplying blood to the sebaceous glands. This kind of context is critical for understanding the role of the vasculature in the ear, as well as improving our understanding of the nuanced interactions of different structures. Figure 4 (b) and (d) show the detailed PARS non-radiative contrast which highlights vasculature throughout the tissue. Similar contrast to previously demonstrated photoacoustic remote sensing systems is achieved providing excellent non-contact non-radiative contrast. Finally, in Figure 4 (c) and (e), the radiative contrast highlights a number of fluorescent biomolecular targets including cartilage matrices, collagen networks, and hair follicles. In addition, the shadows of vasculature can be seen due to the high absorption of hemoglobin to the 532nm light. These key complementary contrasts demonstrate the power of *in vivo* PARS imaging to capture a comprehensive picture of light matter interaction. The value of the exceptional correspondence between contrasts is exemplified through the clear connection of vasculature and underlying structural components of the ear.

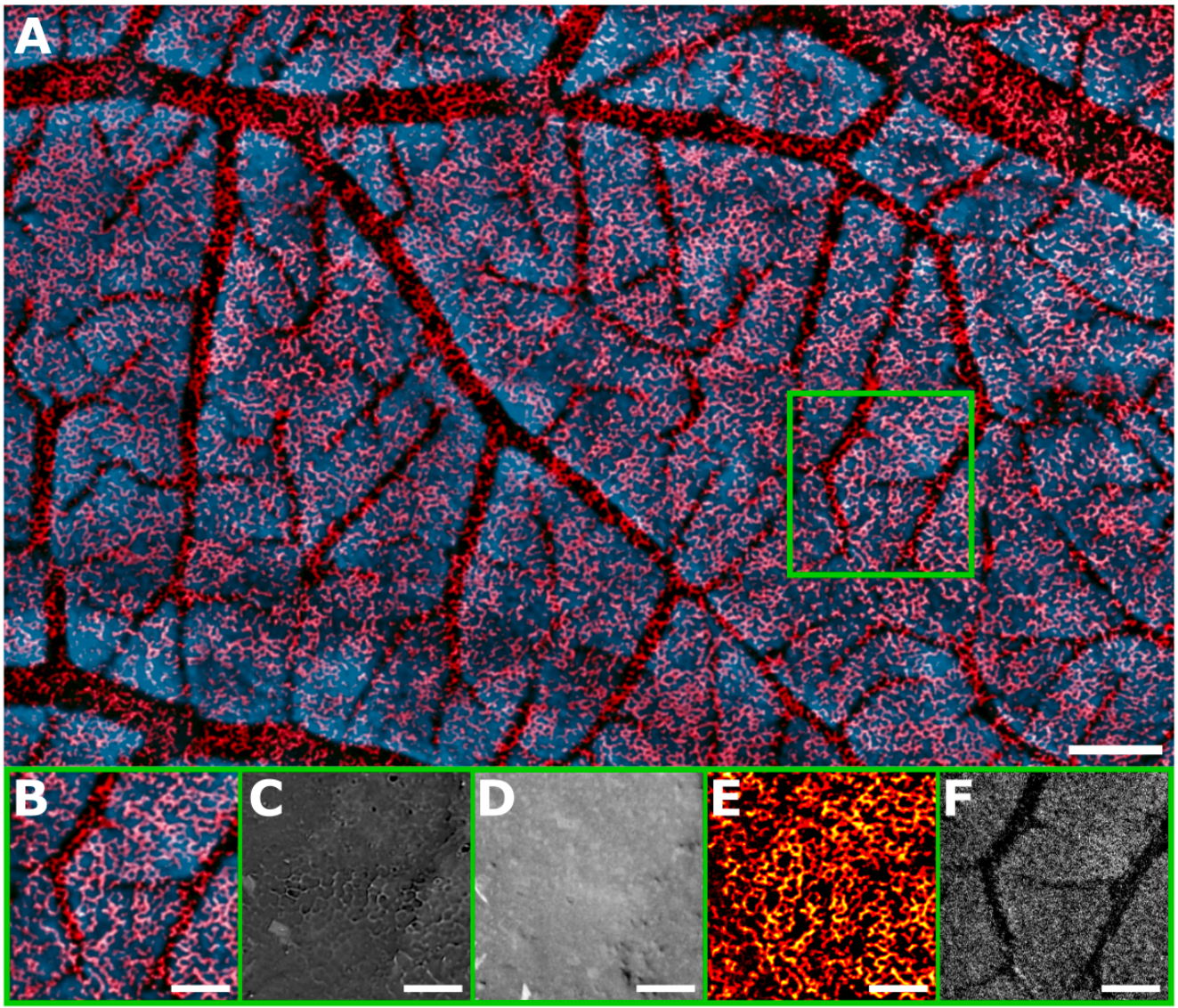

**Figure 5.** PARS images of blood vessels from *in vivo* chicken embryo (CAM) across ~3x2mm. (a) Total absorption PARS image (scale bar: 250μm). (b) Zoom in of total absorption contrast from highlighted area in section A (scale bar: 100μm). (c) Detection scattering contrast from surface morphology. (d) Excitation scattering contrast showing complementary surface morphology contrast. (e) Non-radiative absorption contrast from surface layer capillary vessels. (f) Radiative absorption contrast highlighting deeper vessel structure.

### D. Structural Imaging of Chicken Embryo

The PARS imaging system was used to capture detailed images of the chorioallantoic membrane (CAM) in a chicken embryo model. Figure 5 demonstrates the four captured PARS contrasts from the vasculature in the CAM across ~3x2mm. The full FOV PARS total absorption mosaic can be seen in Figure 5 (a). The TA contrast provides excellent depth

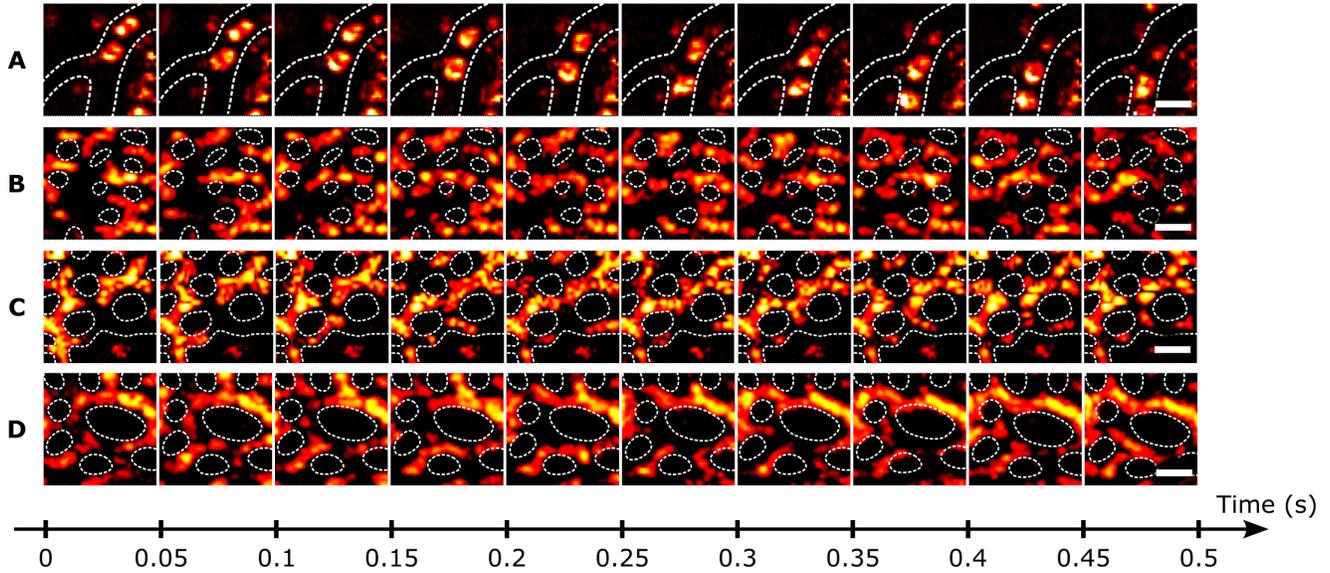

**Figure 6.** Snapshots of PARS non-radiative imaging individual red blood cell flow. The four snapshots (a-d) are from chicken embryo CAM capillary vessel videos captured at 20 FPS for ~20 seconds (scale bar: 20μm). The annotated white dashed lines highlight vessel wall structure alongside the sometimes sparse individual cell motion. See attached supplementary videos for full PARS non-radiative videos captured at 20 FPS

contrast with the surface vessels highlighted in red, and the deeper vessels showing as black shadows against a blue background. Figure 5 (b) further demonstrates this with a zoom in of the highlighted region of CAM in section A. Excellent detail of the actual capillary layer structure can be seen in red with the lower supplying vessels clearly visible in blue. Figure 5 (c) and (d) show the detection and excitation scattering contrast respectively. As can be seen both highlight surface morphology, which captures some of the capillary vessels as they push against the surface. However due to the inconsistent height of the vessels, and additional surface artefacts, an incomplete image of this capillary layer is formed. This highlights the power of the non-radiative PARS contrast shown in Figure 5 (e) which shows great absorption contrast of the capillary layer vessels. Some gaps in contrast are attributed to the dynamically changing red blood cell density within the capillary layer itself which will be explained in the next section. Finally, the radiative PARS contrast shown in Figure 5 (f) shows the shadows of deeper, and larger blood vessels. This is likely due to the large amount of absorbing fluorescent biomolecules sitting deep to the capillary layer vessels. Altogether this demonstrates the power of PARS imaging to capture a motivating image of the entire CAM vasculature network reflection.

### E. Functional Imaging of Blood Flow in Chicken Embryo

Using the faster imaging scan pattern, video rate images of chicken embryo CAM capillary vessels were captured at ~20 frames per second (FPS). Four example video snapshots of the non-radiative PARS contrast are shown in Figure 6 (a-d) with an individual frame size of ~64x64μm. Each snapshot shows ten sequential frames acquired in approximately half a second from each of the different video captures. As can be seen individual red blood cells are fully resolved as they flow through the capillary structures. A variety of sizes and configurations of both vessels and absorbing red blood cells can be seen, with Figure 6 (c) showing the small capillary layer vessels, and Figure 6 (d) highlighting a fork in a larger vessel. Motion of the capillary network structure seen in the videos are attributed to a combination of motion due to the heartbeat of the chicken embryo, as well as bulk movement of the embryo itself. While gaps in the contrast of vessels appears to be due to a combination of red blood cells moving through depth, as well as an often-partial filling of vessel structures themselves. Altogether, the stability PARS non-radiative contrast demonstrated in these videos highlights the power of this *in vivo* PARS microscope as a label-free non-contact functional imaging device. Additionally, the complete resolution of flowing red blood cell structures opens an attractive avenue for future work in extracting accurate blood flow leading to the accurate measurement of the metabolic rate of oxygen consumption using PARS.

Label-Free Non-Contact Structural and Functional Vascular Imaging using Photon Absorption Remote Sensing

## IV. Conclusions

In conclusion, we have demonstrated a novel *in vivo* PARS imaging system which is able to capture key optical interactions within the sample including detection and excitation backscattering, non-radiative relaxation, and radiative relaxation. Additional system developments including significant improvements to imaging performance, stability, and repeatability were achieved through the development of new optical and software systems. System performance was then demonstrated in two established *in vivo* targets, chicken embryo and mouse ear. Where the first videos of individual flowing red blood cells using PARS microscopy were demonstrated. These results are highly motivating for the importance of PARS microscopy as a technique for characterizing tissue structures and function while remaining non-contact and label-free. This work lays the groundwork for PARS imaging in additional targets including the eye and brain where the molecular specificity offered by PARS contrast may be essential. Additionally, this demonstrates feasibility for the accurate measurement of blood flow. Overall, PARS represents an extremely promising technology for the diagnosis and treatment of disease.

**Acknowledgements**
The authors thank Jean Flannigan and all the support staff at the Central Animal Facility at the University of Waterloo for the meticulous care provided to the animals used in this work.

**Funding Sources**
This research was funded by: Natural Sciences and Engineering Research Council of Canada (DGECR-2019-00143, RGPIN2019-06134, DH-2023-00371); Canada Foundation for Innovation (JELF #38000); Mitacs Accelerate (IT13594); University of Waterloo Startup funds; Centre for Bioengineering and Biotechnology (CBB Seed fund); illumiSonics Inc (SRA #083181); New frontiers in research fund – exploration (NFRFE-2019-01012); The Canadian Institutes of Health Research (CIHR PJT 185984).

**Author Contribution Statement**
J.A.T.S. designed and constructed the optical system, created the real time alignment interface, implemented the software processing workflow, conducted the experimental work, prepared the figures, and wrote the manuscript. S.J.W. assisted with all *in vivo* experimental work, design and testing of the digital control system, and experimental design. B.R.E. assisted with preparation of figures, and experimental and optical design. J.T. created stage control software, assisted with preparation of figures, and assisted with experimental design. H.G. maintained all biological specimen labs and organized the acquisition of new animal models. P.H.R conceived the project and supervised the research as the principal investigator and assisted with manuscript writing.

**Competing Interests**
Authors James A. Tummon Simmons, Benjamin R. Ecclestone, James E. D. Tweel, Hager Gaouda, and Parsin Haji Reza all have financial interests in IllumiSonics which has provided funding to the PhotoMedicine Labs. Author Sarah J. Werezak does not have any competing interests.